\documentclass[pre,amssymb,superscriptaddress,floatfix,a4paper,twocolumn,showpacs]{revtex4}
\usepackage{graphicx}
\usepackage{bm}
\usepackage{amsmath}
\usepackage{amsthm}
\usepackage{amssymb}
\usepackage[normalem]{ulem}
\usepackage[usenames]{color}

\newcommand{\fc}{{f_c}}
\newcommand{\fcs}{{f^s_c}}
\newcommand{\vs}{{v^s}}
\newcommand{\zetaRP}{{\zeta^{\tt RP}}}

\begin{document}

\title{
Uniqueness of the thermodynamic limit for driven disordered elastic interfaces
}

\author{A. B. Kolton}
\affiliation{CONICET, Centro At{\'{o}}mico Bariloche, 
8400 San Carlos de Bariloche, R\'{\i}o Negro, Argentina}

\author{S. Bustingorry}
\affiliation{CONICET, Centro At{\'{o}}mico Bariloche, 
8400 San Carlos de Bariloche, R\'{\i}o Negro, Argentina}

\author{E. E. Ferrero}
\altaffiliation[Present address: ]{LiPhy, Universit\'e Joseph Fourier, UMR 5588, 38041 Grenoble Cedex, France}
\affiliation{CONICET, Centro At{\'{o}}mico Bariloche, 
8400 San Carlos de Bariloche, R\'{\i}o Negro, Argentina}

\author{A. Rosso}
\affiliation{LPTMS, Universit\'e Paris-Sud, CNRS (UMR 8626), 91405 Orsay Cedex, France}

\date{\today}

\begin{abstract}
We study the finite size fluctuations at the depinning transition for a one-dimensional elastic interface of size $L$ 
displacing in a disordered medium of transverse size $M=k L^\zeta$ with periodic boundary conditions, where $\zeta$ is the depinning roughness exponent and $k$ is a finite aspect ratio parameter. We focus on the crossover from the infinitely narrow ($k\to 0$) to the infinitely wide ($k\to \infty$) medium. We find that at the thermodynamic limit both the value of the critical force and the precise behavior of the velocity-force characteristics are {\it unique} and $k$-independent. We also show that the finite size fluctuations of the critical force (bias and variance) as well as the  global width  of the interface cross over from a power-law to a logarithm as a function of $k$. Our results are relevant for understanding anisotropic size-effects in force-driven and velocity-driven interfaces.
\end{abstract}


\maketitle
\section{Introduction}
\label{sec:introduction}

Understanding anisotropic finite-size effects in 
driven condensed matter systems is important not only for 
the custom numerical simulation analysis and modeling, 
but also to interpret an increasing  
amount of experiments performed on 
relatively small samples with specially devised geometries,
where one of the system dimensions can be even comparable 
to a typical static or dynamical correlation length.
The steady-state dynamics of directed elastic interfaces in 
random media, experimentally realized in driven
ferromagnetic~\cite{lemerle_domainwall_creep,bauer_deroughening_magnetic2,yamanouchi_creep_ferromagnetic_semiconductor2,metaxas_depinning_thermal_rounding,leekim_creep_corriente}~and ferroelectric~\cite{paruch_ferro_roughness_dipolar,paruch_ferro_quench,JoYang_creep_ferroelectrics,ferroelectric_zombie_paper,paruchbusting}
domain walls, contact lines in 
wetting~\cite{moulinet_distribution_width_contact_line2,frg_exp_contactline}
and fractures~\cite{bouchaud_crack_propagation2,bonamy_crackilng_fracture},  
is a non-trivial relevant example were this kind of phenomenology arises.
The study of driven domain wall motion in 
ferromagnetic micro-tracks for instance~\cite{Parkin_tracks}, relevant for 
memory-device applications or 
metallic ferromagnet spintronics~\cite{ohno2010}, 
motivates the study of their motion in 
``wide'' samples, i.e.
much wider than the interface global width. 
Moreover, at the integration scale for modern nano-devices, 
these tracks can also become thin enough to be 
comparable to the typical size of the thermal nuclei 
controlling creep motion, 
thus yielding an experimentally observable 
dynamical dimensional crossover~\cite{kim2009}.
On the other extreme, periodic systems such as planar vortex 
lattices, charge density waves, or experimental realizations of 
elastic chains in random media, motivate, through an appropriate 
mapping, the study of the motion of 
large interfaces in periodically repeated ``narrow'' media, 
i.e. much narrower than the 
interface width~\cite{cule_1d_elastic_chain_long,bolech_critical_force_distribution,bustingorry_periodic}.

Minimal models, such as the paradigmatic 
quenched-Edwards-Wilkinson equation (QEW) and 
their close quenched disorder variants~\cite{Barabasi-Stanley}, were shown to 
successfully capture experimentally observed 
universal dynamics such as creep~\cite{lemerle_domainwall_creep} and 
depinning ~\cite{frg_exp_contactline,bustingorry_thermal_rounding_fitexp} 
phenomena.
In spite of this, fundamental questions such as 
the possible thermodynamic limits of these models, 
when supplemented with the usual periodic boundary conditions, 
are not completely understood. 
Roughly speaking, the steady-state motion of extended elastic 
interfaces is expected to be very different in very 
narrow than in very wide samples because they actually 
sense rather different pinning force fluctuations from the 
{\it same} microscopic model.
The thermodynamic limit in between these two extremes (i.e., 
not infinitely narrow nor wide samples), hence, looks rather 
ambiguous~\cite{fedorenko_frg_fc_fluctuations}:
it is unclear whether it leads to a unique solution or to a family 
of solutions parametrized by some properly defined 
aspect-ratio parameter.

\begin{figure}[!tbp]
\includegraphics[scale=0.5,clip=true]{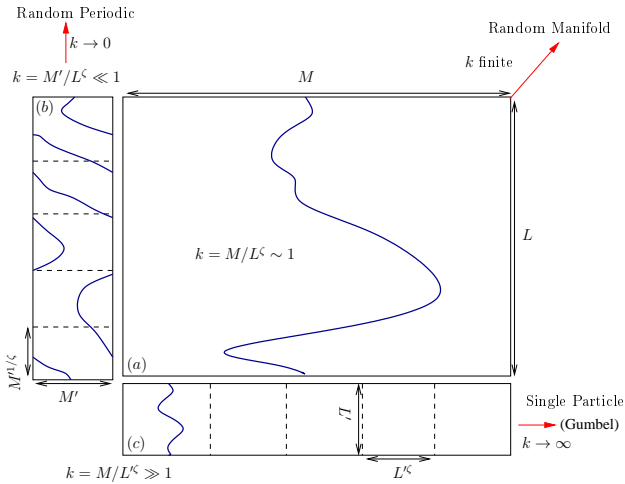}
\caption{\label{fig:schema} (Color online)
Schematic picture showing critical configurations 
for systems with different 
values of the  aspect ratio $k$. 
In (a) the critical configuration  
has a width $w$
comparable to the system transverse dimension $M$. 
In (b), $w \gg M'$ and  
the configuration wraps several times, crossing 
over to the RP geometry, 
$w \sim L^{\zetaRP}$.
In (c) $w \ll M$ and  $w/L^\zeta$ 
is a function of $k$. In the thermodynamic limit, if $k$ is kept constant,
transport properties converge to a unique RM 
limit. If $k \to 0$ the thermodynamic limit corresponds
to the RP class parametrized by the periodicity $M'$.
If $k \to \infty$ the system 
has a dimensional crossover towards the zero-dimensional 
Gumbel class (for Gaussian microscopic disorder), 
and the steady-state motion is static.
}
\end{figure}

Let us consider a driven QEW one-dimensional interface 
in a disordered sample of dimensions $L \times M$, with 
periodic boundary conditions, as schematically shown in Fig.~\ref{fig:schema}(a).
For each sample, at zero temperature, a critical force $\fcs$
separates a pinned  phase from a sliding phase characterized by a finite velocity $\vs$. 
In finite systems, both $\fcs$ and $\vs$ fluctuate from sample to sample and their averages over all disorder realizations, namely
$\langle \fcs \rangle$ and $\langle \vs \rangle$,
depend both on microscopic details of the model
(microscopic disorder distribution, lattice discretization, etc.)
and the specific geometry of the sample
(boundary conditions, transverse size, etc.).
In Ref.~\cite{bolech_critical_force_distribution} it was shown that if we choose 
$M=k L^\zeta$, with $\zeta$ the depinning roughness exponent, 
and a Gaussian microscopic disorder, the critical force 
distribution crosses over 
from a Gaussian (for $k\to 0$) to a Gumbel distribution 
(for $k\to \infty$) in the large $L$ limit. 
One can show that the ``infinitely narrow'' $k\to 0$ limit  corresponds 
to the so-called random periodic (RP) depinning universality class, while the ``infinitely wide'' $k \to \infty$ 
limit corresponds to a dimensional crossover towards 
the zero dimensional case describing a single particle in an effective 
one-dimensional potential. 
In the former case, periodic effects arise when $M$ turns out to be comparable
to the interface width, and becomes more and more important as $k$ decreases 
further and the interface winds more around the cylinder with perimeter $M$, 
as schematically shown in Fig.~\ref{fig:schema}(b).
In the latter case instead, as schematically shown in 
Fig.~\ref{fig:schema}(c), periodicity effects are absolutely negligible
but the roughness turns anomalous due to extreme value statistics 
effects~\cite{bustingorry_paperinphysics}.
In fact, for fixed $L$ and Gaussian disorder,
$\langle \fcs \rangle \to \infty$ as $M\to \infty$; so that, 
at zero temperature, a finite velocity is only possible 
in a non-steady state~\cite{ledoussal_driven_particle}.
The so-called random manifold (RM) regime, 
which exists between these two extreme limits for any finite $k$, 
was shown to display a {\it finite} critical force $\fc$ 
in the thermodynamic limit ($\fc = \lim_{L \to \infty} \langle \fcs \rangle$), 
with sample to sample fluctuations vanishing as~\cite{bolech_critical_force_distribution,middleton_cdw,narayan_fisher_cdw} 
$\langle[\fcs-\langle \fcs \rangle]^2\rangle 
\sim L^{-2/\nu} =L^{-2(2-\zeta)}$,  given the STS relation $\nu=1/(2-\zeta)$.

The previuos results represent a considerable 
progress in the understanding of the finite-size effects 
but still leave us with a rather ambiguous 
picture for applications and analytical calculations. 
Important open questions are: 
(i) What is the dependence of $\fc$, and $\langle[\fcs-\langle \fcs \rangle]^2\rangle$ 
with the self-affine aspect-ratio parameter $k$?
(ii) Is there a finite-size bias 
$[\fc-\langle \fcs \rangle]$ and how does it depend on $L$ and $k$?. 
(iii) Is the RM thermodynamic limit prescription for the 
critical force, $\lim_{L,M=k L^\zeta \to \infty}$, different from the 
one for extracting the RM velocity-force 
characteristics? 
In the affirmative case, how does it depend on $k$?
(iv)  Geometry and transport are closely related, as changes in the interface 
velocity directly affect the location of one or 
several geometrical crossovers~\cite{kolton_depinning_zerot2,kolton_dep_zeroT_long,ferrero_comptes_rendus}.
How sensible to the value of $k$ are the geometry and the velocity of the interface?
In this paper we address these open questions and show that 
constant force simulations in finite samples 
actually lead to an unambiguous thermodynamic 
critical force and velocity-force characteristics, which is 
independent of $k$, as long as $k$ is finite.
We also show how the finite system transport properties scale 
towards 
the (RP) $k\to 0$ and (single particle or extreme RM) 
$k\to \infty$ limits as a function of $L$.
Finally we discuss how our results relate to the cases of velocity-driven 
interfaces and other alternative methods used to define the 
thermodynamic critical force.

\section{Model, Observables and Method}

We consider the  driven QEW model at zero temperature, described by 
\begin{equation}\label{eq:eqmotion}
\gamma \partial_t u(x,t) = c \partial_x^2 u(x,t) + F_p(u,x) + f.
\end{equation}
This equation models the overdamped dynamics of the displacement field $u(x,t)$
of a one-dimensional elastic interface in a two-dimensional random medium.  
We will consider here  a sample of size $L \times M$ with periodic boundary conditions in both directions.
The pinning force derives from a bounded random potential 
(i.e. Random Bond disorder), $F_p(u,x) = -\partial_u U(u,x)$, 
with correlations
\begin{equation}
\label{eq:correlator}
\langle{[U(u,x) - U(u',x')]^2}\rangle= R(u-u')\delta(x-x'),  
\end{equation}
with $R(y)$ a short-ranged function of range $r_f$ and $\langle \cdots \rangle$ standing for the average over all disorder realizations.

In particular we study a model of Eq.~\eqref{eq:eqmotion} where the displacement field is discrete in the 
$x$-direction, and  the random potential $U(u,x)$ is given by a sequence of uncorrelated random (Gaussian) numbers glued by a piece-wise cubic-spline with $r_f=1$. The details of the model are  described elsewhere~\cite{rosso_depinning_simulation}.  
For each sample there is a {\em unique} critical force $\fcs$ and a {\em unique} critical configuration $u^s_c(x)$.
Both quantities  are computed 
in an efficient and accurate way without actually 
solving the true dynamics~\cite{rosso_dep_exponent,rosso_depinning_simulation}. 
For $f>\fcs$, at long times, the interface acquires a steady state velocity $\vs$.
To obtain it, we solve the dynamics of Eq.~\eqref{eq:eqmotion} 
from an arbitrary initial condition up to very long times using 
a parallel algorithm~\cite{ferrero_nonsteady}.
We define the width $w$ of the critical configuration:
\begin{equation}
w^2(L,M) = \left\langle \frac{1}{L} \sum_{x=0}^{L-1} [u^s_c(x)-{u_{cm}}^s_c]^2 \right\rangle,
\label{eq:roughness-def}
\end{equation}
with ${u_{cm}}^s_c=(1/L) \sum_{x=0}^{L-1} u^s_c(x)$ the sample dependent center of mass position of the critical configuration.
When $M\sim L^\zeta$ it is well known that 
$w^2 \sim L^{2\zeta}$ with~\cite{ferrero_nonsteady} $\zeta=1.250 \pm 0.005$. 

In the following, we analyse the large size limit of (i) the critical force, (ii) the velocity for a fixed value of the external force and (iii) the width of the critical configuration,
as a function of both $L$ and $k$, from the infinitely narrow sample to the infinitely wide sample. 

\section{Results}
\label{sec:results}

\subsection{Summary of Finite-Size Scaling Results}
Here we summarize our main results.
Note that in one dimension, the roughness exponent for the RM class
is~\cite{ferrero_nonsteady} $\zeta = 1.250$, and for the RP class
is~\cite{ledoussal_frg_twoloops,bustingorry_periodic} $\zetaRP=1.5$.
A detailed description and discussion of each 
result is left for the next sections. The main results are:
\begin{enumerate}
  \item  The critical force reaches a $k${\it -independent} value 
  $\fc = \lim_{L,M \to \infty} \langle \fcs \rangle$ in 
  the thermodynamic limit $L,M \to \infty$ for any {\it finite} $k=M/L^\zeta$ 
  (see Fig. \ref{fig:Fcthermolimit}). 
  This value only depends on microscopic details of the system.
  The velocity-force characteristics also displays the same convergence 
  towards a unique $k$-independent thermodynamic value, 
  $v(f)=\lim_{L,M \to \infty} \langle v^s(f) \rangle$.
  (see Fig.~\ref{fig:Vthermolimit}). 

  \item The average finite-size critical force $\langle \fcs \rangle$ 
  approaches the value $\fc$ from above if $k$ is large, 
  and from below if $k$ is small, compared to a (non-universal) marginal 
  value $k^* \approx 2.1 \pm 0.1$ 
  (see Figs. \ref{fig:Fcthermolimit} and \ref{fig:Fcthermolimit2}).
  The asymptotic forms for this bias are well described by 
  \begin{equation}\label{eq:fcbias}
  (\fc-\langle \fcs \rangle) L^{2-\zeta} \sim 
  \left\{ 
  \begin{array}{rl}
   k^{1-2/\zeta} &\mbox{ if $k \ll k^*$} \\
   -(\log k)^{1/\delta} &\mbox{ if $k \gg k^*$}
  \end{array} 
  \right.  
  \end{equation}
  with $1<\delta<2$ (see Figs.~\ref{fig:Fcbiassmallk} and \ref{fig:Fcbiaslargek}), 
and consistent with the critical force 
distribution tail of the form $\ln P(\fcs,L,M=k^* L) 
\sim -\fcs^\delta$ with $\delta=1.2 \pm 0.1$. (see Fig.~\ref{fig:fcdistribution}).

  \item The sample to sample fluctuations of the critical force are
  well described by  
  \begin{equation}\label{eq:fcfluctuations}
      (\langle \fcs^2 \rangle -\langle \fcs \rangle^2) L^{2(2-\zeta)} \sim 
  \left\{ 
  \begin{array}{rl}
   k^{1-\zetaRP/\zeta} &\mbox{ if $k \ll k^*$} \\
   (\log k)^{-2(1-1/\delta)} &\mbox{ if $k \gg k^*$}
  \end{array} 
  \right.  
  \end{equation}
  so they decrease with increasing $L$ or $k$
  (see Figs.~\ref{fig:fcfluctuationssmallk} and \ref{fig:fcfluctuationslargek}).

  \item The width of the critical configuration behaves as 
  \begin{equation}\label{eq:roughness}
      w L^{-\zeta} \sim 
  \left\{ 
  \begin{array}{rl}
   k^{-(\zetaRP/\zeta-1)} &\mbox{ if $k \ll k^*$} \\
   (\log k)^{\zeta/2\delta} &\mbox{ if $k \gg k^*$}
  \end{array} 
  \right.  
  \end{equation}
  The roughness thus always increases with increasing $L$ but has a 
  non-monotonous behaviour with $k$, decreasing for small $k$ and 
  increasing for large $k$.
  Its minimum is reached for a value $k \lesssim k^*$.
  (see Fig.~\ref{fig:roughness}).

\end{enumerate}

\subsection{The thermodynamic limit of the critical force}

\begin{figure}[!tbp]
\includegraphics[scale=0.3,clip=true]{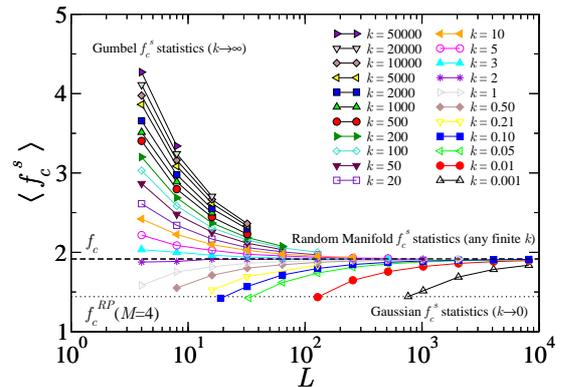}
\caption{(Color online)
Longitudinal finite-size dependence of the averaged critical 
force for samples of size $L\times M$, with $M=k L^\zeta$.
The (non-universal) thermodynamic limit $\fc$ attracts 
any finite value of $k$, thus determining this property 
unambiguously for the RM class. Also indicated is the 
(non-universal) thermodynamic RP critical 
force for a periodicity $M=4$. Note that for $k=k^* \approx 2$, 
finite-size effects are negligible for several decades of $L$.  
} \label{fig:Fcthermolimit}
\end{figure}

\begin{figure}[!tbp]
\includegraphics[scale=0.3,clip=true]{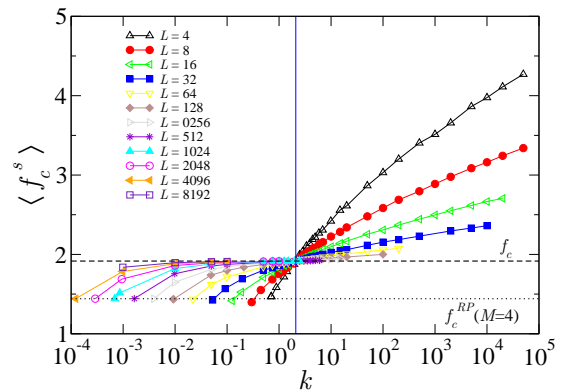}
\caption{(Color online)
Same data as in Fig.\ref{fig:Fcthermolimit} but showing the 
transverse finite-size dependence, for each $L$, 
of the averaged critical force.
Note the crossing at $k^*\approx 2$ (precisely we get
$k^*=2.1 \pm 0.1$), bridging 
the Gaussian ($k\to 0$) with the Gumbel ($k\to \infty$) 
critical force statistics, where 
finite-size effects become negligible.
} \label{fig:Fcthermolimit2}
\end{figure}
 
The existence of a {\it unique} critical force $\fc$ and 
velocity-force characteristics $v(f)$ (see Sec.~\ref{sec:velocityfinitesize}) 
for all finite values of $k$ shows that
transport properties of the QEW model have an 
unambiguous thermodynamic limit.  
In other words, the infinite family of systems described 
by (the same) Eq.~(\ref{eq:eqmotion}) but with (different)
geometries parametrized by the self-affine aspect-ratio parameter
$k$, is attracted to a unique RM behaviour in the 
large size limit (see Fig. \ref{fig:Fcthermolimit}).
Note also that, even if $\fc$ and $v(f)$ 
are not universal, they are intrinsic; i.e., they only depend on the 
parameters appearing in the microscopic equation of motion.

\subsubsection{Finite-Size effects for small $k$}

\begin{figure}[!tbp]
\includegraphics[scale=0.3,clip=true]{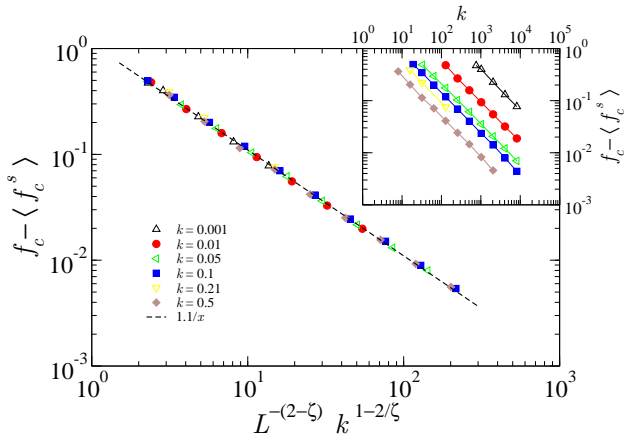}
\caption{(Color online)
Negative finite-size bias of the critical force, $\fc - \langle \fcs\rangle$, for 
small $k$ values. Inset: raw data. Main: scaled data. In this regime 
the shift is controlled exclusively by $M$ and by
the RM depinning roughness exponent $\zeta$.
} \label{fig:Fcbiassmallk}
\end{figure}

\begin{figure}[!tbp]
\includegraphics[scale=0.3,clip=true]{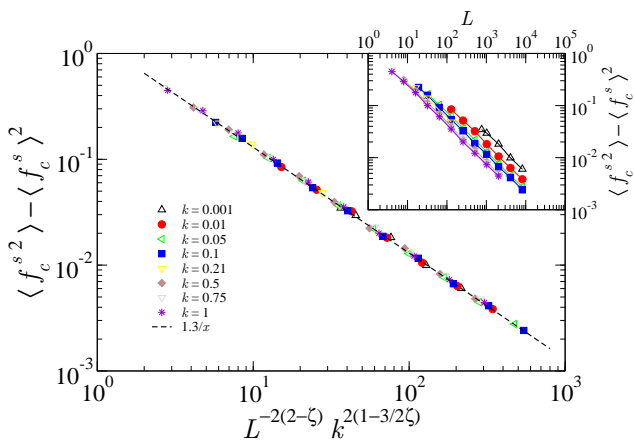}
\caption{(Color online)
Finite-size fluctuations of the sample critical force 
$\langle \fcs^2 \rangle - \langle \fcs\rangle^2$, for 
small $k$ values. Inset: raw data. Main: scaled data. In this regime 
the fluctuations are well described by 
$\langle \fcs^2 \rangle - \langle \fcs\rangle^2 \sim L^{-2(2-\zeta)} k^{2(1-3/2\zeta)}$.
Note that in order to have a non-zero  
$\lim_{k\to 0} [\langle \fcs^2 \rangle - \langle \fcs\rangle^2]$, we need 
$\zeta \to 3/2$ in agreement to what 
corresponds to the (one-dimensional) RP class 
$[\langle \fcs^2 \rangle - \langle \fcs\rangle^2] \sim L^{-1}$ 
and roughness exponent $\zetaRP=3/2$. 
} \label{fig:fcfluctuationssmallk}
\end{figure}

When $k$ is much smaller than $k^* \sim \mathcal{O}(1)$ (i.e. $M \ll L^{\zeta}$), 
the system can wrap around $M$ several times and, thus, 
sense the transverse periodicity of the disorder. 
Indeed, at a characteristic length $L_M \sim M^{1/\zeta} \ll L$
the geometry of the interface crosses over from the RM
to the RP class, parametrized by the 
periodicity $M$. 
This crossover is well manifested in the structure factor 
$\langle |u_c(q)|^2 \rangle$ of the critical 
configuration, which displays~\cite{bustingorry_periodic}, at $q\sim L_M^{-1}$, 
a crossover
from the RM roughness exponent $\zeta=1.250$ 
(i.e., $\langle |u_c(q)|^2 \rangle \sim 1/q^{1+2\zeta}$) 
, to the RP exponent
\footnote{Note that $\zetaRP \equiv \zeta_{\tt L} = (4-d)/2$ in $d=1$,
where $\zeta_{\tt L}$ is the Larkin exponent.
This reflects the fact that the RP
fixed point with $\zeta=0$ is unstable 
due to a self-generated random-force~\cite{ledoussal_frg_twoloops}.}
$\zetaRP=3/2$ (i.e., $\langle |u_c(q)|^2 \rangle \sim 1/q^{1+2\zetaRP}$) 
increasing the observation lengthscale $q^{-1}$.

When $L$ grows with fixed $k$, $L_M$ grows as $L_M \sim k^{1/\zeta} L$.
Since the average critical force $\langle \fcs \rangle$ and velocity 
$\langle v^s \rangle$ for a finite system are determined by the typical 
behaviour at small length-scales ($l<L_M$) or short wavelength modes, 
the critical force must approach the RM thermodynamic 
value $\fc$ as $L_M$ grows, 
even when the large scale geometry ($l>L_M$) is still described by
$\zetaRP$ instead of $\zeta$.
Since for our model of Eq.~(\ref{eq:eqmotion}), the RP critical 
force $f^{\tt RP}_c \simeq \lim_{L\to \infty} \langle \fcs(L, M) \rangle$ 
with $M \sim 1$ is always smaller that the RM critical 
force $\fc$, for small $k$ the thermodynamic limit is 
approached from \emph{below}, 
as can be observed in Figs.~\ref{fig:Fcthermolimit} and~\ref{fig:Fcthermolimit2}.
Furthermore, we can see in Fig.~\ref{fig:Fcbiassmallk} a 
negative finite-size bias of the critical force $\fc-\langle \fcs \rangle$
for small values of $k$ that smoothly follows Eq.~\eqref{eq:fcbias}.

Let us introduce an heuristic argument to understand the scaling.
In principle, one can think the string as being composed 
by $L/L_M = k^{-1/\zeta} \gg 1$ ``RM blocks'', of 
longitudinal size $L_M$ and transverse size $M$. 
Note that each of these blocks has precisely the ``proper'' 
aspect-ratio $L_M^\zeta/M =1$.
If we consider that each of these blocks participates in
the total critical force with independent
contributions, such that they average to $\overline{f_M}$ with a dispersion $\sigma_M$,
where $\overline{\cdots}$ stands for an average over the independent RM blocks, then we can write:
\begin{eqnarray}
\langle \fcs(L,M) \rangle \approx \overline{f_M} \\
\langle \fcs^2(L,M) \rangle-\langle \fcs(L,M) \rangle^2 \sim
\frac{\sigma_M^2}
{{L/L_M}},    
\end{eqnarray}
The above assumptions are consistent with the 
fact that $\fcs$ has almost a  Gaussian statistics~\cite{bolech_critical_force_distribution}
if $M \ll L^\zeta$, by virtue of the central limit theorem for the sum of 
many pinning forces with finite variance, which are 
uncorrelated at distances smaller than $L_M$.
If $\overline{f_M}$ represents minus the average pinning force
on a given block of size $L_M \times M$, then we can write
$\overline{f_M} \sim \fcs(L_M,M)$. 
Since the interface in each block is, by definition, in the RM regime, 
we can write $\sigma_M \sim L_M^{\zeta-2}$ for its sample to sample
fluctuations. 
We thus get:
\begin{eqnarray}
\langle \fcs(L,M) \rangle \approx \langle  \fcs(L_M,M) \rangle \\
\langle \fcs^2(L,M) \rangle-\langle \fcs(L,M) \rangle^2 \sim {k^{2(1-3/2 \zeta)}} L^{-2(2-\zeta)} 
\end{eqnarray}
First, let us note that the predicted finite-size scaling 
dependence on $L$ and $k$ for $\langle \fcs^2(L,M) \rangle-\langle \fcs(L,M) \rangle^2$ 
is indeed what we observe in Fig.~\ref{fig:fcfluctuationssmallk}.
Second, let us note that $\langle \fcs \rangle$ depends only 
on $M$ (as $L_M = M^{1/\zeta}$), in consistency with the behaviour shown 
in Fig~\ref{fig:Fcbiassmallk}.
If this bias scales with the longitudinal size in the same way as the 
sample to sample fluctuations, one can predict 
$[\fc - \langle \fcs(L,M) \rangle] \approx [\fc - \langle  \fcs(L_M,M) \rangle] \sim L_M^{\zeta-2} \sim M^{2/\zeta-1} \sim L^{-(2-\zeta)}k^{(1-2/\zeta)}$, 
as shown in Fig.~\ref{fig:Fcbiassmallk}.

It is interesting to note that in order to have  
$\lim_{k\to 0} \left[\langle \fcs^2(L,M) \rangle-\langle \fcs(L,M) \rangle^2\right] 
\sim {k^{2(1-3/2\zeta)}} L^{-2(2-\zeta)}$ finite in the $L\to \infty$ thermodynamic limit,
which corresponds to the RP class with a fixed periodicity $M$, 
we require that $\zeta \to \zetaRP=3/2$.
Therefore, 
$\langle \fcs^2(L,M) \rangle-\langle \fcs(L,M) \rangle^2 \sim L^{-1}$, 
again consistent with the prediction for the critical force fluctuations
in the RP class.

\subsubsection{Finite-Size effects for large $k$}

\begin{figure}[!tbp]
\includegraphics[width=8cm,clip=true]{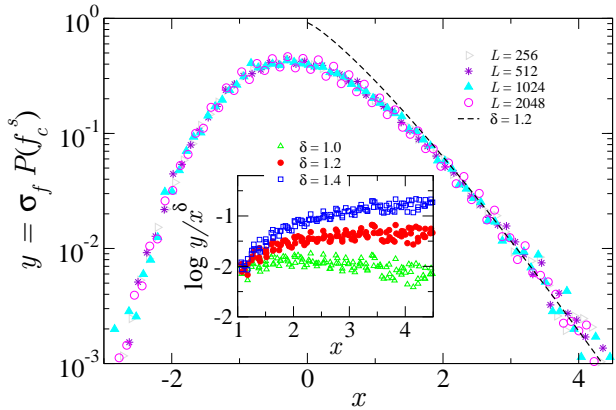}
\caption{ \label{fig:fcdistribution} (Color online) Scaled distribution function for the critical force, $y= \sigma_f P[f^s_c(k=2)]$ against $x=  (\fcs - \langle \fcs \rangle)/\sigma_f$, with $\sigma_f=\sqrt{(\langle \fcs^2 \rangle - \langle \fcs\rangle^2)}$. Different system sizes are shown, $L \times M$ with $M=k L^\zeta$ for $k=2$. The large $f^s_c$ tail can be fitted to a stretched exponential, $P(f^s_c) \sim \exp[-{f^s_c}^{\delta}]$, characterized by $\delta = 1.2 \pm 0.1$, as can be observed in the inset.
}  
\end{figure}

When $k$ is large, periodicity effects disappear, since the critical 
configuration cannot wind around the cylinder of perimeter $M$.
In turn, we start to observe {\it extreme value statistics} effects.
As described in~\cite{bolech_critical_force_distribution}, 
in the $k\to \infty$ limit the critical force distribution tends to a Gumbel 
function; 
i.e., the critical force of each sample can be thought as the maximum of
independent identically
distributed (iid) variables with a (stretched) exponential-tailed distribution.
This explains why $\langle \fcs(L,M) \rangle$ approaches $\fc$ 
from above in Fig.~\ref{fig:Fcthermolimit}.
Indeed, considering $\langle \fcs \rangle$ as the maximum among 
the critical forces of many independent configurations
we can expect a growth with increasing $M$ at $L$ fixed,
which is observed in Fig.~\ref{fig:Fcthermolimit2}.
Since metastable (or quasi-critical) configurations just below (or 
above) the depinning transition are essentially decorrelated in 
a distance of the same order than its width 
$w \simeq L^\zeta$, the number of such independent 
random variables is precisely $k=M/L^\zeta$.

Let us analyse the case when $k$ is finite and close to $k^*$. On one hand, in this case finite size 
effects are less pronounced, as shown in Fig.~\ref{fig:Fcthermolimit}. 
We can expect that, in the same sense that $\fcs$ is attracted to $f_c$, 
the distribution function should also be attracted to a thermodynamic limit which would 
be close to the one with $k=k^*$.
Note that the critical force $\fcs$ of a system 
of size $L \times k^* L^\zeta$ is distributed according to a 
function which is intermediate between 
the Gaussian and the Gumbel's~\cite{bolech_critical_force_distribution}. 
Although when the shape of this function 
is not known analytically, 
we know that it must decay 
faster than a power law, since the maximum among $k=M/L^\zeta$ 
of such systems is attracted to the Gumbel class in the 
$k\to \infty$ limit.
To support this idea, let us consider the tail of $P(\fcs)$
described in particular as a stretched exponential decay, 
$\ln P(\fcs)\sim -{\fcs}^\delta$, with 
$1 \leq \delta \leq 2$ (the bounds corresponding  
to the Gaussian case, $\delta=2$,  and to the 
Gumbel case, $\delta=1$).
In order to test this idea, we show
in Fig.~\ref{fig:fcdistribution} the distribution function for the sample critical
force $\fcs$ corresponding to $k=2$ where the system size effects are almost negligible.
It is shown in this case that the tail exponent characterizing the stretched exponential behaviour is
$\delta = 1.2 \pm 0.1$.

\begin{figure}[!tbp]
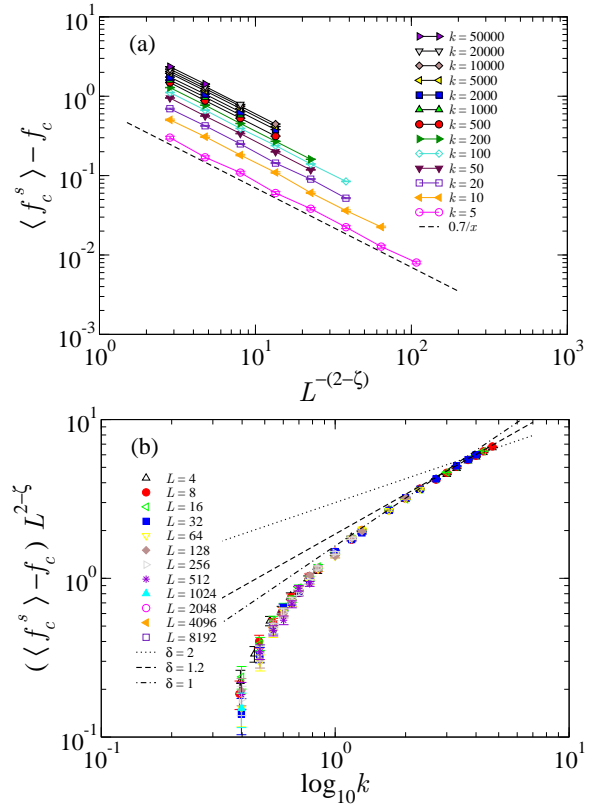

\includegraphics[scale=0.3,clip=true]{figures/DfcvsL.eps}
\includegraphics[scale=0.3,clip=true]{figures/Dfcvsk.eps}
\caption{\label{fig:fig4} (Color online)
Positive finite-size shift of the critical force, $\fc - \langle \fcs\rangle$, for 
large $k$.
(a) The shift decays as $L^{-(2-\zeta)}$, with a $k$-dependent prefactor. 
(b) As expected from extreme value statistics,  at the largest $k$ the shift increases with $k$ logarithmically, $\fc - \langle \fcs\rangle \sim (\log k)^{1/\delta}$, with $\delta = 1.2$ in agreement with the tail exponent obtained in Fig.~\ref{fig:fcdistribution}. We compare with $\delta=2$ 
expected for the maximum of iid Gaussian random variables and $\delta=1$ for for the maximum of iid exponential 
variables.
} \label{fig:Fcbiaslargek}
\end{figure}

\begin{figure}[!tbp]
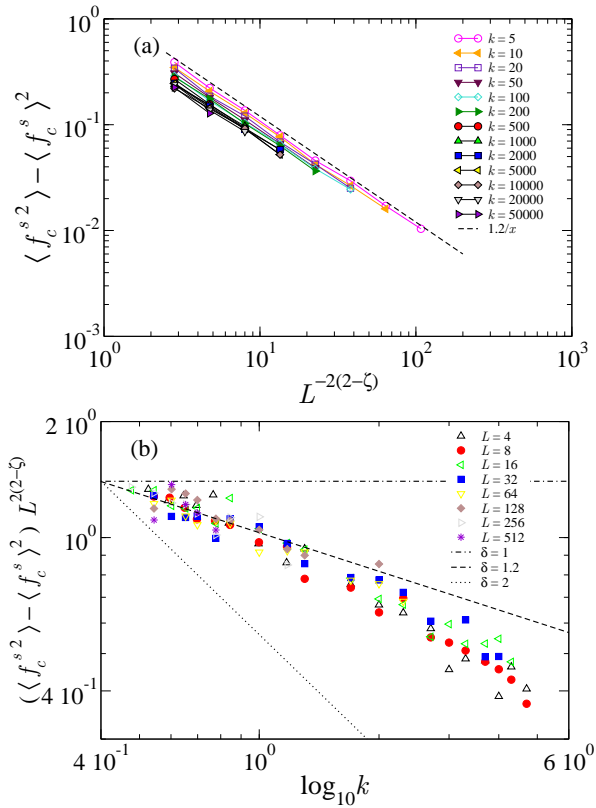

\includegraphics[scale=0.3,clip=true]{figures/dfcvsL-largek.eps}
\includegraphics[scale=0.3,clip=true]{figures/dfcvsk-largek.eps}
\caption{(Color online)
Finite-size fluctuations of the critical force 
$\langle \fcs^2 \rangle - \langle \fcs\rangle^2$, for 
large $k$.
(a) $\langle \fcs^2 \rangle - \langle \fcs\rangle^2 \sim L^{-2(2-\zeta)}$, 
same as in the low $k$ regime (Fig.~\ref{fig:fcfluctuationssmallk}). 
(b) Dependence with $k$. We find 
$\langle \fcs^2 \rangle - \langle \fcs\rangle^2 \sim (\log k)^{-2(1-1/\delta)}$, 
consistent with the finite-size shift expected from extreme statistics. Note 
the consistency with $\delta=1.2$ from Fig.~\ref{fig:Fcbiaslargek}. 
} \label{fig:fcfluctuationslargek}
\end{figure}

Therefore, based on the observed stretched exponential behaviour and from standard
extreme value statistics arguments~\cite{Galambos} we get that 
the average of $\fcs = \max\{f^{(0)}_1,f^{(1)}_1, ..., f^{(k)}_1\}$ should 
grow as $\langle \fcs \rangle - \fc \sim L^{-(2-\zeta)}(\log k)^{1/\delta}$.
Here we have used again that the finite size 
bias for a $L \times L^\zeta$ (or $k\sim 1$) system behaves 
as the sample to sample fluctuations 
$\langle f_1 \rangle - \fc \sim L^{-(2-\zeta)}$ 
in the $L \to \infty$ limit.
This prediction is consistent to what 
is found in the simulations, as shown in Fig.~\ref{fig:Fcbiaslargek}.
The large $k$ behaviour is consistent with the obtained value of the tail exponent $\delta = 1.2$.
Within this picture, standard extreme value arguments also predict 
$\langle f^2_c \rangle -\langle \fcs \rangle^2 \sim L^{-2(2-\zeta)} (\log k)^{-2(1-1/\delta)}$. 
This is also consistent with Fig.~\ref{fig:fcfluctuationslargek}, 
where we compare the prediction using the same value of $\delta$ obtained 
in Fig.\ref{fig:fcdistribution}.
As it can be observed in Fig.~\ref{fig:fcfluctuationslargek}(b), in the large $L$ limit the data is consistent with the value of
$\delta = 1.2 \pm 0.1$, ruling out the bounds $\delta=2$ for the maximum of iid 
Gaussian variables, and $\delta=1$ for the maximum of 
iid exponential variables.

It is interesting to note that since $\delta>1$ the sample to sample 
fluctuations of the critical force {\it decrease} for increasing $k$,
unlike the mean value of the critical force, which {\it increases} with $k$.
Therefore, the growing sample critical force reaches a sharply 
defined value in the large $k$ limit.
This might be important for experiments.

\subsection{Finite-size effects in the roughness of the critical configuration}

\begin{figure}[!tbp]
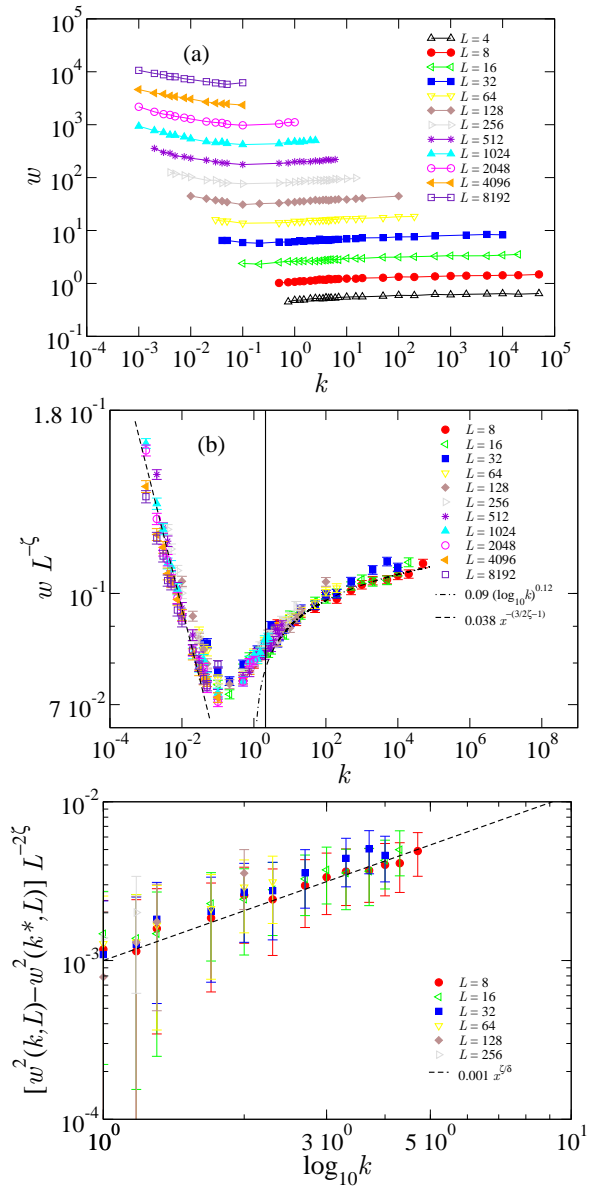

\includegraphics[scale=0.3,clip=true]{figures/wvsk.eps}
\includegraphics[scale=0.3,clip=true]{figures/wLzvsk.eps}
\includegraphics[scale=0.3,clip=true]{figures/wLzvsk-largek-b.eps}
\caption{\label{fig:fig9} (Color online)
Anisotropic finite-size analysis of the width $w$ 
of the critical configurations. (a) Raw-data. 
(b) Scaled data, according to $w \sim L^\zeta$, vs $k$. 
In the low $k$ regime, $w \sim L^\zeta k^{-(\zetaRP/\zeta-1)}$ (dashed-line). 
Note that for a non-zero $\lim_{k\to 0} w$ we need $\zeta \to \zetaRP=3/2$, corresponding to the 
(one-dimensional) RP class 
(Compare with the critical-force fluctuations in the same regime shown 
in Fig.~\ref{fig:Fcbiassmallk}). 
The width for large $k$ is roughly described by some power-law of 
$\log k$ (dot-dashed line).
The solid line indicates $k^*=2.1$.
(c) Scaled data, according to $[w^2(k,L)-w^2(k^*,L)] 
\sim L^{2\zeta} (\log k)^{\zeta/\delta}$ and using the values $\zeta = 1.250$ and $\delta = 1.2$, showing the agreement between the data and the scaling prediction.
} \label{fig:roughness}
\end{figure}

The typical interface global width or roughness also manifests 
size effects at depinning, which are consistent with the 
finite-size scaling for the critical force.

Let us first 
describe the low-$k$ behaviour of the width $w$ shown in Fig~\ref{fig:roughness}.
From the study of the RM to RP crossover 
at $L_M \sim M^{1/\zeta} \ll L$ we know
that~\cite{bustingorry_periodic} $w(L) \sim L_M^\zeta (L/L_M)^{\zetaRP}$.
We thus get
\begin{equation}
 w(L) \sim k^{-(\zetaRP/\zeta-1)} L^{\zeta}  
\end{equation}
in consistency with the asymptotic behaviour shown in Fig.~\ref{fig:roughness}(b) 
for small values of $k$, where $\zetaRP=3/2$ for our one-dimensional case.

Turning now to the behaviour of the roughness critical configurations 
at large values of $k$, we can see that, interestingly, 
it displays an approximate logarithmic growth with 
$k$, as shown in Fig.\ref{fig:roughness}(b).
This was already observed 
in Ref.~\cite{bustingorry_paperinphysics}. Here we link 
such behaviour with the critical force statistics and predict 
its scaling form.
We start by noting that in order to have a logarithmic growth of 
$\fcs$ with $k$ in this regime, either the individual pinning 
forces on the monomers of the critical configuration get more 
correlated in order to increase $\fcs$, or they remain uncorrelated 
but acquire an enhancement of their dispersion with increasing $k$.
Since we do not observe increased correlations between 
the individual pinning forces acting on the monomers of 
$u_c$ for large $k$, the last scenario 
is the most plausible.
A logarithmic enhancement in the prefactor $w^2/L^\zeta$ 
can be thus heuristically understood as follows. 
From the Larkin formulation we can define effective Larkin length 
$L_c \sim (\fc/c r_f)^{-1/2}$ and roughness $w \approx r_f (L/L_c)^{\zeta}$.
Extending this idea to sample to sample fluctuations, we consider that
$\langle L_c^s(k,L) \rangle \sim (\langle \fcs \rangle/c r_f)^{-1/2}$ and 
$w \approx r_f (L/\langle L_c^s \rangle)^{\zeta}$.
Therefore, using that $\langle \fcs \rangle - \fc \sim (\log k)^{1/\delta}$
we easily get $w(k,L)/L^{\zeta} \sim (\log k)^{\zeta/2\delta}$ 
in the very large $k$ limit, where $\langle \fcs \rangle \gg \fc$. 
Since our data does not reach such limit, we can not neglect the $\fc$ 
contribution. 
A corrected version reads
$(w(k,L)/L^{\zeta})^2 - (w(k^*,L)/L^{\zeta})^2 \sim (\log k)^{\zeta/\delta}$,
where we have used the $k$-independent thermodynamic limit
$\lim_{L\to \infty} w(k,L)/L^{\zeta} \approx w(k^*,L)/L^{\zeta} \sim r_f L_c^{-\zeta}$.
Fig.~\ref{fig:roughness}(c) presents a scaled version of the data to test this idea, where we have used that $\zeta = 1.250$, $\delta = 1.2$ (see fit in Fig.~\ref{fig:Fcbiaslargek}) and $k^* \approx 2$
(see Fig.~\ref{fig:Fcthermolimit2}), and shows a very good agreement with the scaling prediction.

In summary, the increasingly rare critical configurations 
for large $k$ can be thus seen as being pinned 
by an effectively stronger uncorrelated microscopic disorder
keeping the same elasticity and microscopic disorder correlator range. 
In other words, extreme statistics
shorten the effective Larkin length on those critical 
configurations.

\subsection{Finite-size effects in the velocity-force characteristics}
\label{sec:velocityfinitesize}

\begin{figure}[!tbp]
\includegraphics[scale=0.3,clip=true]{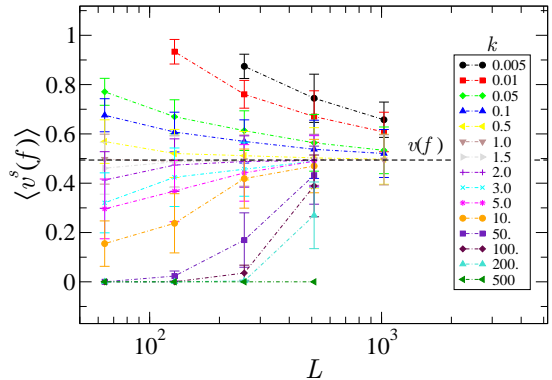}
\caption{(Color online)
Finite size effects in the mean velocity for different values of 
the self-affine aspect ratio $k$, and the same driving force $f=1.95 > \fc$. 
We observe that velocity curves for finite $k$ values are attracted to the same limit $v(f) \approx 0.5$, 
except for $k = 500$, where rare blocking configurations still dominate and 
$\fcs > f$ for the range of $L$ values analysed.
The black dashed line indicates the estimation of the stationary velocity 
in the thermodynamic limit $\lim_{L\to \infty} \langle v^s(f) \rangle|_k \to v(f)$.
} \label{fig:Vthermolimit}
\end{figure}

\begin{figure}[!tbp]
\includegraphics[scale=0.7,clip=true]{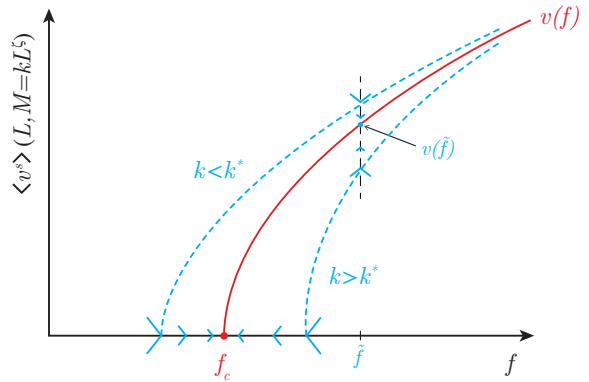}
\caption{(Color online)
Schematic $k$-dependence of the velocity-force characteristics. For $k>k^*$ the critical force bias is positive, $\fcs < f_c$, and therefore the velocity at a given force $f>f_c$ approaches the thermodynamic limit from below, $\langle v^s(f) \rangle < v(f)$. The opposite is observed when $k<k^*$: the critical force bias is negative and hence the velocity is $\langle v^s(f) \rangle > v(f)$.
} \label{fig:vfscheme}
\end{figure}

\begin{figure}[!tbp]
\includegraphics[scale=0.3,clip=true]{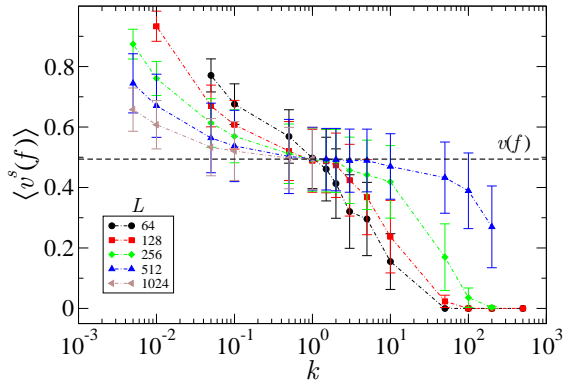}
\caption{(Color online)
Same data as in Fig.~\ref{fig:Vthermolimit} but showing the transverse
finite-size dependence, for each L, of the averaged stationary velocity.
Note the crossing at $k^*_v$, where finite-size effects become negligible, 
bridging the Gaussian ($k\to 0$) with the Gumbel ($k\to \infty$) statistics.
} \label{fig:Vthermolimit_k}
\end{figure}

When $f>\fcs$, the elastic interface moves steadily with velocity 
$v^s(f)$ and the geometry displays a crossover in the roughness from 
the exponent $\zeta$ to the exponent $\zeta^{\tt th}=(2-d)/2$ 
at the characteristic scale $\xi \sim v^{-\nu/\beta}$.
Given the presence of this extra length-scale depending on the 
force excess, it is not obvious whether the same thermodynamic 
limit prescription for the RM critical force
(i.e., to fix the aspect-ratio parameter $k = ML^{-\zeta}$)
will work for the velocity, yielding a unique 
$k$-independent velocity limit $v(f)$.

In Fig.~\ref{fig:Vthermolimit} we show that the prescription 
for $\fc$ works well for $v(f)$. 
We observe that a wide range of values of $k$ tends to
converge to a $k$-independent, force dependent steady-state velocity, 
$\lim_{L\to \infty} \langle v^s(f;k,L) \rangle|_k \to v(f)$.
At finite $L$, we observe that $\langle v^s(f) \rangle > v(f)$ 
for small $k$, and $\langle v^s(f) \rangle < v(f)$, for large $k$.
This is consistent with the behaviour of the finite-size average 
critical force (see schema in Fig.~\ref{fig:vfscheme}):
$\fcs$ is biased to greater values of $f$
as $k$ increases above $k^*$, so, if we assume a monotonous and
continuous behaviour of the velocity-force curve, for a given $f$
the velocity average $\langle v^s(f) \rangle$ should be smaller
as $k>k^*$ increases.
For very large $k$ indeed, we can be in the situation where $f<\fcs$
and thus $\langle v^s \rangle=0$, as is seen in 
Fig.~\ref{fig:Vthermolimit} for $k=500$.
On the other hand, for small values of $k<k^*$, $\fcs$ decreases with $k$,
and at a fix force $f$ the average velocity will be larger as $k$ decreases.
This is why curves for $k<k^*_v\simeq 1 \sim k^*$ converge from \textit{above} in
Fig.~\ref{fig:Vthermolimit} to the thermodynamic limit.

In Fig.~\ref{fig:Vthermolimit_k} we observe the behavior of
$\left<v^s(f)\right>$ as a function of $k$ for different system sizes.
For the working force $f=1.95>\fc$, a crossing of all curves at 
$k^*_v\simeq 1$ can be observed,
but also, how curves are apparentely atracted to a unique constant value,
both avobe and below $k^*_v$, as $L$ increases.
We find that $k^*_v \sim \mathcal{O}(1)$, 
without appreciably varying with $f$. 
This shows that the same critical force prescription  
is adequate to obtain the thermodynamic limit of the 
velocity-force curve, and that finite 
size effects at any finite $k$, vanish as $L\to \infty$.

\section{Discussion}

\subsection{Comparison with the velocity-driven ensemble}
\label{sec:velocitydriven}

In this paper we have defined the critical force of a one-dimensional QEW line in a given 
finite disorder sample of dimensions $L \times M$ with periodic boundary conditions 
in all directions, driven by a uniform, constant force $f$.
In some situations however, the interface 
is velocity-driven in a infinitely wide medium, and the driving force $f$ is replaced 
by a term $m^2 [vt - u(x,t)]$, with $v$ the imposed 
mean velocity.
We will argue that the results of Eqs.~\eqref{eq:fcbias},
and \eqref{eq:fcfluctuations} are also relevant for this case, 
and that a close connection exists simply by relating the curvature parameter 
$m$ and the transverse periodic dimension $M$.
Since the parabolic drive sets a characteristic 
length-scale $L_m \sim 1/m$ in the longitudinal direction, 
we can compare it directly with the length $L_M \sim M^{1/\zeta}$  
set by the periodic boundary conditions in the constant force simulations. 
We can hence relate $M^{1/\zeta}$ and $1/m$, so the 
limit of small $m$ corresponds to the large $M$ limit, and 
viceversa.

The critical force is defined 
in the quasistatic limit of the velocity-driven interface as 
$\langle \fcs(L,m) \rangle \equiv \lim_{v\to 0+} m^2 \langle [vt -u(x,t)] \rangle$
for stationary values of $u(x,t)$
It can be compared with the critical force 
$\langle \fcs(L,M) \rangle$ discussed in the previous subsections. 
Functional Renormalization group (FRG)
calculations predict when $L \to \infty$
that $\langle \fcs(m) \rangle= \fc + c_1 m^{2-\zeta}$ in the small $m$ limit, 
with $c_1$ a negative constant and $\fc$ the thermodynamic 
critical force. 
If we assume $L$ very large and 
define $k_m = (L m)^{-\zeta}$, such prediction reads 
$\langle \fcs(m) \rangle= \fc + c_1 k_m^{1-2/\zeta} L^{\zeta-2}$. 
As shown in Eq.~\ref{eq:fcbias} this 
is exactly the same scaling we find for 
$\langle \fcs(L,M=kL^\zeta) \rangle$ for small $k$, with 
$\fc > \langle \fcs(m) \rangle$ assured by the FRG prediction $c_1<0$. 
This supports our identification of $k$ with $k_m$, and we
can expect Eqs.~\eqref{eq:fcbias} and
\eqref{eq:fcfluctuations} to hold in the velocity-driven 
ensemble by replacing $k \mapsto k_m$.

To further emphasize the connection we note that 
the numerical extrapolation of $\langle \fcs(m) \rangle$ 
in the velocity-driven ensemble yield a value~\cite{rosso_correlator_RB_RF_depinning}
$\fc \sim 1.9$, 
indistinguishable from ours, $\fc \approx 1.916 \pm 0.001$, for the 
same microscopic disorder.
On the other hand, the prediction
$\langle \fcs(m) \rangle= \fc + c_1 k_m^{1-2/\zeta} L^{\zeta-2}$ 
shows that for small $k_m$ (i.e. large $m$ compared 
to $L^{-1}$), the critical force is smaller than $\fc$, 
as we see in Fig.~\ref{fig:Fcbiassmallk} for small $k$. 
Moreover, as shown in Fig.~2 of Ref.~\cite{rosso_correlator_RB_RF_depinning},
when $k_m$ is large (i.e., $m$ small compared to $L^{-1}$), 
$\fcs(m)$ can become larger than the extrapolated $\fc$.
This is due to extreme value statistic effects 
similar to the ones discussed in the previous sections: 
as the curvature of the parabola vanishes for a fixed 
$L$, the interface can get blocked in more rare 
configurations with systematically higher critical forces. 

In summary, the transport properties have a unique limit 
and similar finite-size effects in the two ensembles.
Only the roughness of the critical configurations for 
small $k$ or $k_m$ are different, since for the velocity-driven case, the 
roughness beyond the length-scale $L_m\sim 1/m$ crosses over 
from $\zeta \approx 1.25$, to $\zeta^{\tt m}=0$ 
(instead of $\zetaRP=3/2$), so $w \sim m^{-\zeta}$.
On the other hand, for small $m$, such that 
$mL \ll 1$, we expect to observe 
a behaviour analogous to Eq.~\eqref{eq:roughness} for 
large $k$.
This has not been analyzed yet in the velocity-driven ensemble.

Finally, it is worth mentioning that an original 
alternative approach was analytically implemented 
in Ref.~\cite{fedorenko_frg_fc_fluctuations}, by defining 
a critical force for a fixed center of mass position.
This choice avoids rare configurations as the interface can not 
explore the disorder in the transverse directions beyond 
the length-scale set by its own width $w\sim L^\zeta$. 
It is thus equivalent to work with a system-size satisfying 
$k \sim 1$ (or $k_m\sim 1$), and must thus have the same 
unique thermodynamic limit.
The advantage of this method is that it is parameter free, and 
size effects are only controlled by $L$.
In addition, it does not present crossovers, and the critical 
configuration geometry always belong to the (non-extreme)
RM class. This method has been so far 
only implemented analytically however.

\subsection{Implications for the non-steady universal relaxation of the velocity}
\label{sec:relationnonsteady}

It is interesting to relate the finite-size bias of the critical 
force, Eq.~\eqref{eq:fcbias}, with the universal non-steady 
relaxation at the thermodynamic depinning threshold~\cite{kolton_short_time_exponents}. 
In Ref.~\cite{ferrero_nonsteady} it was noted that the short-time relaxation 
of an initially flat interface {\it at the RM thermodynamic critical 
force} $\fc$ can be effectively described as an interface 
of ``size'' $\ell(t)$ which is quasistatically driven by the finite-size bias 
of the critical-force. That is, we assume that $v(t)$ 
instantaneously satisfies the steady-state relation 
$v(t) \sim [f_c - f_c(\ell(t))]^{\beta}$, where the 
effective ``size'' grows with time as the growing correlation
length $\ell(t)\sim t^{1/z}$, with $z$ the dynamical exponent. 
By assuming $f_c - f_c(\ell(t)) \sim \ell(t)^{\zeta-2}$, we get 
the critical relaxation $v(t) \sim t^{-\beta/\nu z}$.

The finite size scaling of Eq.~\eqref{eq:fcbias} allows us 
now to better justify the above assumptions. 
The initially flat relaxing string of size $L$, 
in the small (non-steady) velocity limit such that the 
adiabatic approximation holds, 
effectively becomes a (pseudo) critical configuration confined in a system 
of effective size $L \times w(t)$. 
This situation 
is equivalent to the one described 
in Sec.\ref{sec:velocitydriven} with 
$m(t) \sim w(t)^{-1/\zeta}$, in the quasistatic drive 
limit. This defines an effective 
aspect-ratio parameter $k(t) \sim w(t)/L^\zeta \ll k^*$, 
and allow us to write, 
\begin{eqnarray}
v(t) \sim [f_c - f_c(L,k(t)))]^{\beta} \sim k(t)^{\beta(1-2/\zeta)} \sim t^{-\beta/\nu z}, 
\end{eqnarray} 
where in the second term we have used the $k \ll k^*$ scaling for the 
bias of the critical force, Eq.~\eqref{eq:fcbias}, and in the third term the STS relation $\nu=1/(2-\zeta)$.
When $k(t) \sim k^*$ the bias vanishes, corresponding to the vanishing 
of the velocity in a finite system when~\cite{kolton_short_time_exponents}
$\ell(t)\sim L$. 
The string is then blocked by a {\it typical} RM critical configuration. 
In order to explore rare critical configurations we need to drive the system above 
the thermodynamic critical force $f > \fc$. Then, 
from Eq.~\eqref{eq:fcbias}, and the same adiabatic approximation 
for $v(t)$, we can expect a crossover to a new 
regime in the non-steady relaxation, from a 
power-law to a slower logarithmic decay.

\section{Conclusions} 

We have shown that there exists a unique, unambiguous thermodynamic limit 
for the transport properties of driven elastic interfaces 
in random media, irrespective of the precise 
relation between the longitudinal and transverse
dimensions of the system, only provided they maintain a 
definite scaling relation in the large size limit. Namely, 
any finite value of the self-affine aspect-ratio parameter 
$k = M/L^\zeta$, with $\zeta \simeq 1.250$ the depinning exponent, 
leads to exactly the same transport properties in the 
large-size limit.  
We have also characterized in details the finite-size effects 
in the critical force fluctuations 
for small and large values of $k$. Our results thus extends the one of
Ref.\cite{bolech_critical_force_distribution} in several useful ways. 
In particular, we show that the thermodynamic critical force is not only finite if 
$k$ is finite, but that it is independent of $k$; i.e., it is \textit{unique}.
We also report a finite-size bias or shift in the critical force, 
which was unnoticed before, as, in general, only reduced variables 
(of zero mean) were analyzed.
We give good evidences that the velocity-force characteristics is itself
a \textit{unique} curve in the thermodynamic limit, and interpret it as an 
attractor for the stationary and even non-stationary behavior of any finite
system with well defined geometry.
Finally, we have also shown that our results are completely consistent with 
the ones obtained for velocity-driven interfaces, where $f \to m^2(vt - u(x,t))$,
so the two ensembles have the same transport properties in the thermodynamic limit
and are thus equivalent. 

\section*{Acknowledgements}
A.B.K. and A.R. acknowledge partial support by the France-Argentina
MINCYT-ECOS A12E05.
E.E.F. acknowledges partial support by the France-Argentina 
Bernardo Houssay Program 2012.
S.B. acknowledges partial support by the France-Argentina
MINCYT-ECOS A12E03.
Partial support from Project PIP11220090100051 (CONICET)
and Project PICT2010/889 are also acknowledged.

\bibliography{tfinita5}

\end{document}